\documentclass[journal,twoside,shortpaper,9pt]{IEEEtran}

\usepackage{textcomp}
\usepackage[noadjust]{cite}
\usepackage{amsmath,amssymb,amsfonts}
\usepackage{algorithmic}
\usepackage{graphicx}
\usepackage{multicol,multirow}
\usepackage{array}
\usepackage{epstopdf}
\usepackage[utf8]{inputenc}
\usepackage{color}

\newcolumntype{L}{>{\raggedright\arraybackslash}p}
\newcolumntype{C}{>{\centering\arraybackslash}m}

\begin{document}

\title{A Broadband Cavity-Backed Slot Radiating Element in Transmission Configuration}
\author{Alberto Hernández-Escobar, Elena Abdo-Sánchez, \textit{Member, IEEE,} and Carlos Camacho-Peñalosa, \textit{Senior Member, IEEE}

\medskip
\textit{(DOI: 10.1109/TAP.2018.2874069) © 2018 IEEE*}
\thanks{* Personal use of this material is permitted. Permission from IEEE must be obtained for all other uses, in any current or future media, including reprinting/republishing this material for advertising or promotional purposes, creating new collective works, for resale or redistribution to servers or lists, or reuse of any copyrighted component of this work in other works.

Manuscript received July 27, 2017. This work was supported in part by the European Union’s Horizon 2020 research and innovation programme under the Marie Sklodowska-Curie grant agreement No. 706334, in part by the Spanish Ministerio de Educación, Cultura y Deporte (Programa para la Formación del Profesorado Universitario) under Grant FPU15/06457 and in other part by the Spanish Ministerio de Economía y Competitividad, under the project ADDMATE TEC2016-76070-CR3-3-R.

The authors are with the Departamento de Ingeniería de Comunicaciones, Escuela Técnica Superior de Ingeniería de Telecomunicación, Universidad de Málaga, Andalucía Tech, 29010 Málaga, Spain (e-mail: ahe@ic.uma.es).}}


\maketitle

\begin{abstract}
\boldmath
A planar technology stripline-fed slot radiating element in transmission configuration is proposed. Its main advantages are the broad impedance bandwidth achieved and the property that it only radiates into half-space, which are obtained, respectively, with the use of its complementary strip element and using a cavity-backed slot. A lattice network circuit model is proposed both to explain the behavior of the structure and to establish a design methodology. Its capabilities are shown through simulation and demonstrated in a proof of concept prototype. Measurement results show a unidirectional broadside radiation pattern and a fractional bandwidth of 48\%, significantly superior to other slot-based radiating elements found in the literature. The element has the ideal characteristics for building series-fed reconfigurable arrays for wide-band applications.

\end{abstract}

\begin{IEEEkeywords}

Broadband radiating element, cavity-backed slot, lattice network, leaky-wave antennas, stripline.

\end{IEEEkeywords}

\section{Introduction}
\label{sec:introduction}

Society's increasing needs for wide-bandwidth wireless communications demands broadband, directive and reconfigurable antennas to be easily integrated into small terminals. Slot-like antennas have been studied and used intensively in recent decades [1] due to their many advantages, which include low cost, low profile, durability, easy manufacture and integration in the casing of almost any device. Therefore they are strong candidates for building arrays that can comply with the aforementioned requirements.

Nevertheless, these antennas usually radiate towards the entire space, thus exhibiting bilateral radiation. This feature makes them poorly suited for their use in conventional directive arrays since the radiation pattern will always have at least two main lobes. In order to prevent radiation towards one of the half-spaces, a cavity made of a conductive material can be placed behind the slot, forming a so-called Cavity-Backed Slot (CBS). This structure was studied and used during the second half of the last century [2], [3] founding that the radiation properties are determined by the dimensions of the cavity. Later, these antennas were left out due to their bulkiness and complex manufacture using planar technology. The development of Substrate Integrated Waveguide (SIW) technology opened up a way for a compact and easy implementation of CBS antennas [4], [5] by using metallic via holes to build the cavity.

In addition to the aforementioned drawbacks, slot and CBS antennas have a narrow impedance bandwidth due to the resonant nature of the element. A recent work [6] uses the SIW technique and a bow-tie slot antenna to increase its impedance bandwidth, but still only 9.4\% is achieved. A solution to the narrow bandwidth which uses transmission configuration and offers an ultra broad bandwidth was proposed in [7] for the case of microstrip-fed slots. Very broad bandwidth is achieved by using a stub complementary to the slot that matches the impedance of the structure, resulting in an all-pass section. The two-port configuration of this radiating element enables the design of tunable series-fed arrays, as shown in [8]. However, the bilateral radiation problem still exists for the so-called \textit{complementary strip-slot}. In an attempt to address this issue, a reflector was placed behind the array in [8] and [9] with the drawbacks of narrowing the working band of the antenna and considerably increasing its size.

A novel compact stripline-fed CBS radiating element using planar technology is proposed in this communication. In a similar way as in [7], its complementary stub is placed under the slot to enhance the impedance bandwidth. The two-port transmission configuration is also adopted here and the cavity is implemented using SIW technology. The present work can be seen as a transformation of the \textit{complementary strip-slot} to obtain unidirectional radiation. This is achieved by replacing the microstrip line by stripline, preventing the structure from radiating towards one half-space. These modifications in the geometry entail several challenges: the CBS behaves differently than that of a conventional slot, the stripline makes it more difficult to obtain impedance matching, and controlling the modes supported by the structure becomes a fundamental part of the design.

Stripline-fed CBSs were also researched in the last century [10]--[12] and were proposed [11] as a candidate for series-fed arrays. More recently, the stripline-feeding approach has been used as an alternative to cavity feeding [13]. Furthermore, they have previously been used in conjunction with SIW technology [14]. The impedance bandwidth of the CBS antennas found in the literature is usually very narrow, less than 5\%, for both one-port and two-port (transmission) configurations. The design presented in this manuscript shows a very broad fractional impedance bandwidth (48\%). Thus, to the authors' knowledge, the proposed radiating element exhibits the highest impedance bandwidth from among the other unidirectional radiating narrow slots cited in the bibliography. With these characteristics, the element can be used to make wide-band series-fed antenna arrays as in [8], but featuring unidirectional radiation.

The manuscript is structured as follows: Section II introduces the antenna geometry and the modes supported by the structure. Section III extracts an equivalent circuit based on the lattice network. Section IV describes the design methodology used. Section V presents a design in the 5 GHz band and provides simulation and measurement results. Section VI discusses several aspects about the use of the element in series-fed arrays. Finally, the conclusions are summarized in Section VII.

\section{Structure and Supported Modes}

\begin{figure}[t!]
\centering
\centerline{\includegraphics[width=0.9\columnwidth]{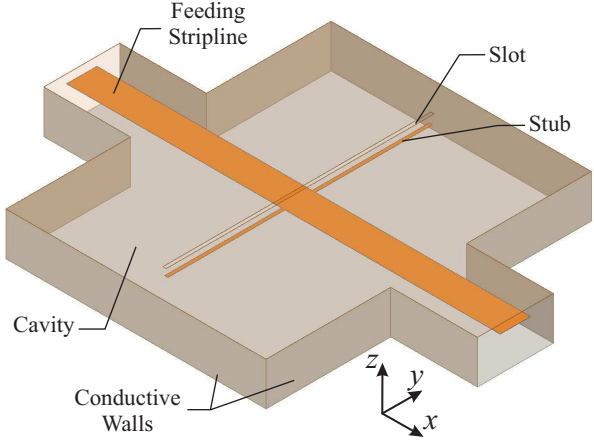}}
\caption{Geometry of the proposed structure.}
\end{figure}

The proposed structure consists of a CBS excited in transmission configuration by an asymmetric stripline where a stub, complementary to the slot, is placed beneath it. The slot is etched on the upper ground plane and the stub is aligned with it but placed on the stripline layer. Lateral metallic walls are used to keep the structure closed, building the cavity, which is filled with dielectric material. Fig. 1 shows the geometry of the proposed radiating element.

\begin{figure}[t!]
\centering
\centerline{\includegraphics[width=0.95\columnwidth]{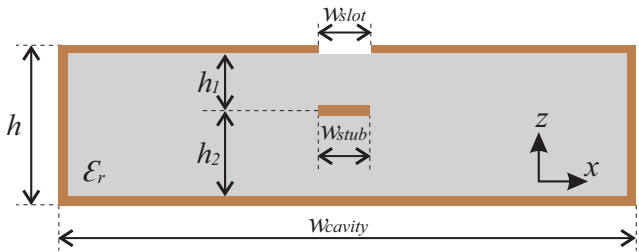}}
\caption{Cross-section of the simulated transmission structure used to compute the modes supported by the structure.}
\end{figure}

The behavior of the structure is determined by the modes supported by the section of the structure (shown in Fig. 2) along the direction transverse to the feeding line ($y$-axis in Fig. 1). The fundamental mode of this structure is a quasi-TEM mode with no cut-off, supported by the two conductors in the structure (strip and cavity walls), in a similar way to a classic stripline TEM mode. Due to the slot, the structure is not completely closed as is a classic stripline. However, the field leaked outside the structure by the slot is not significant and, thus, it is possible to assume that the fundamental mode is almost a TEM mode. Another mode supported by the structure is the slot mode of a CBS. Although there are some similarities between the field distribution of this mode and that of the slot mode of a classic slotline, the dimensions of the cavity play a fundamental role in its behavior. Furthermore, as this mode is not affected by the strip, it is supported by a single-conductor structure (the cavity walls) and, thus, it has a cut-off frequency. Furthermore, higher-order modes can also appear. The first higher order mode supported by this structure is the TE$_{10}$ mode of the rectangular waveguide made up of the metallic walls and the upper and the lower ground planes of the cavity.

\begin{figure}[t!]
\centering
\centerline{\includegraphics[width=\columnwidth]{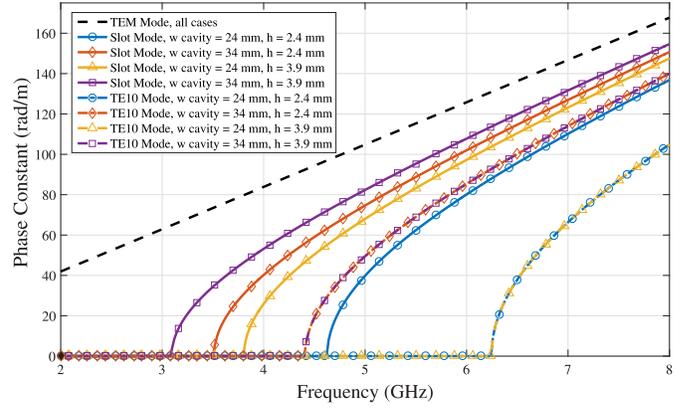}}
\caption{Simulated phase constants of the three first modes propagating through the transmission structure for four different cavity cross-sections.}
\end{figure}

As will be justified in Section III, in order to cancel out the resonant behavior of the slot and achieve broad matching, it is necessary for both the slot and the quasi-TEM modes to be excited. If the dimensions of the cavity are chosen properly, the element can successfully operate in a wide frequency band. The structure's lowest working frequency is limited by the cut-off frequency of the slot mode of the CBS, because its propagation is necessary for the structure to radiate. The TE$_{10}$ mode is unwanted and its appearance will limit the working frequency band at higher frequencies. The cut-off frequency of the unwanted TE$_{10}$ mode could limit the highest working frequency. Thus, its cut-off frequency should be as high as possible. However, since the other modes inside the structure barely excite the TE$_{10}$ mode, the presence of this mode may go unnoticed even above its cut-off frequency. Furthermore, since the transmission system is short-circuited at its ends (by the metallic walls that close the structure), a cavity is formed. The TE$_{10}$ mode will resonate inside the cavity at the resonance frequency of the resonant mode TE$_{101}$, as explained in [15]. The effect of this higher order mode will become noticeable at this resonant frequency.

A simulation of the transmission system made up of the cavity, shown in Fig. 2, has been carried out for different dimensions using the ANSYS HFSS commercial electromagnetic simulator. Fig. 3 displays the phase constants of the first three modes of this transmission system to illustrate how these modes are supported by the structure and how their cut-off frequencies change as the cavity dimensions are modified. The dielectric used in this simulation is air and the width of the slot is 0.3 mm. To increase the cut-off frequency of the TE$_{10}$ mode, a narrow cavity, with a low $w_{cavity}$, is preferred. However, if the cavity width, $w_{cavity}$, is reduced, the cut-off frequency of the slot mode will increase. To broaden the operating frequency band, the cut-off frequency of the slot mode can be lowered by increasing the height of the stripline and thus the cavity, $h$. However, the radiating element will be thicker, as extracted from the study of Fig. 3. The width of the slot, $w_{slot}$, also changes the cut-off frequency of its mode. Lower values of the width reduce the frequency and thus are desirable. However, very low values of $w_{slot}$ could be difficult to implement accurately and could reduce the radiation of the element significantly. Furthermore, to increase the bandwidth where the TE$_{10}$ mode is supported but not excited, a high resonance frequency of the resonant mode TE$_{101}$ is needed. To do this, the cavity must be as short as possible, with a low $l_{cavity}$. However, the length of the cavity is limited by the length of the slot.

In conclusion, the dimensions of the cavity must be chosen carefully due to the trade-off between the operating bandwidth of the element and its thickness. Results from Fig. 3 show that, when $w_{cavity}$ is 24 mm and $h$ is 3.9 mm, there is a frequency band between 3.8 GHz and 6.2 GHz where only the desired modes, stripline and slot, are supported. If the TE$_{10}$ mode is not excited beyond 6.2 GHz and if the length of the cavity is 35 mm, the working frequency band would extend up to 7.5 GHz, which corresponds to the resonance frequency of the resonant TE$_{101}$ mode in this case.

\begin{figure}[t!]
\centering
\centerline{\includegraphics[width=0.8\columnwidth]{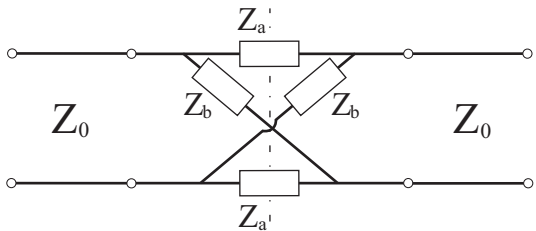}}
\caption{Equivalent lattice network of the complementary structure.}
\end{figure}

\begin{figure}[t!]
\centering
\centerline{\includegraphics[width=\columnwidth]{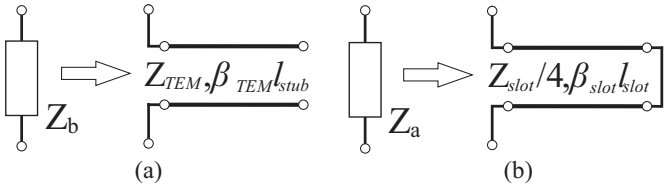}}
\caption{Circuit model of the lattice network impedances using transmission lines. (a) $Z_b$. (b) $Z_a$.}
\end{figure}

\section{Lattice-Network-Based Equivalent Circuit}

In an analogous way as in [7], a lattice network ([16], Fig. 4) is proposed to model the element. The field distribution of a symmetric structure can be separated into two contributions: the field distribution of the even mode and the field distribution of the odd mode. Thus, if the reference planes are chosen appropriately (coincident at the symmetry plane, in this case), the lattice network separates the contribution of both the even and odd modes in each of the branches, allowing the creation a model of the structure from independent models of the even and odd modes. These two modes correspond to the TEM mode of the stub and the slot mode of the CBS as described in the previous section.

Fig. 5 shows the circuit model of the impedances of each branch. The major differences with the model proposed in [7] are that, in the case of $Z_b$, the impedance of a stripline, $Z_{TEM}$, (which can be extracted from [17]) and the propagation constant of a pure TEM mode, $\beta_{TEM}$, are used instead. In the case of $Z_a$, its modeling is more challenging than in [7] since analytical expressions to obtain the impedance of the slot mode, $Z_{slot}$, and its phase constant, $\beta_{slot}$, have not been found in the literature. For this reason, the propagation constant of a classic slotline and the simulation results of the impedance of the slot are used as an approximation.

To improve the accuracy of the model, a lossy transmission line which would take into account the radiation of the slot, can be used. The attenuation constant of the line, $\alpha_{slot}$, has to be obtained from the simulation of the complete structure. This is due to the fact that the radiation of the slot heavily depends on the coupling between the feeding stripline and the slot. Therefore, both the width of the feeding line and its distance to the slot play a fundamental role.

\section{Design}

Firstly, the dimensions of the cross-section of the cavity ($w_{cavity}$, $h$, $\varepsilon_r$, and $w_{slot}$ in Fig. 2) must be chosen according to the following two criteria: first, obtaining a cut-off frequency of the slot mode lower than the design frequency and, second, achieving a cut-off frequency of the TE$_{10}$ as high as possible. The effect of these dimensions on the cut-off frequencies has already been discussed in Section II.

Secondly, the length of the slot, $l_{slot}$, is chosen so that the resonance of the slot coincides with the design frequency. The slot resonance is modeled with a transmission line as shown in Fig. 5(b). As the propagation constant, $\beta_{slot}$, is already determined in the previous step, the length of the slot, $l_{slot}$ will only modify the electrical length of the transmission line. This length will also determine the minimum length of the cavity, $l_{cavity}$, which sets the resonance of the resonant mode TE$_{101}$.

Lastly, the other parameters, $w_{stub}$, $l_{stub}$ and $h_1$ must be chosen in order to obtain broad impedance matching. To do this, first, let us express the image impedance of the equivalent circuit of the complementary structure shown in Fig. 4 as a function of $Z_a$ and $Z_b$, as in [16]:
\begin{align}
\label{eq1}
Z_{im}(\omega)=\sqrt{Z_a(\omega)Z_b(\omega)}.
\end{align} The image impedance of the structure must be constant and with the same value as the characteristic impedance of the feeding line, $Z_0$, over a wide range of frequencies. As modeled in Fig. 5, the impedances $Z_a$ and $Z_b$ have poles and zeros at the corresponding resonance frequencies. For $Z_{im}$ to be approximately constant, these poles and zeros must coincide in frequency so they properly cancel each other out and the impedance level does not change over frequency. For this to happen, the electrical length of the transmission lines modeling $Z_a$ and $Z_b$ must be the same. These conditions can be expressed as follows:
\begin{subequations}
\begin{align}
\label{eq2}
\frac{1}{2}\sqrt{Z_{TEM}Z_{slot}}=Z_0
\end{align}
\begin{align}
\label{eq3}
\beta_{TEM} l_{stub} = \beta_{slot} l_{slot}.
\end{align}
\end{subequations} Although it is possible to modify $Z_{TEM}$ using the width of the stub, $w_{stub}$, this may not be enough to fulfill condition (2a), since stripline structures present lower impedances than their microstrip counterpart. Increasing $h$ allows a higher stripline impedance, and thus, a higher $Z_{TEM}$ to be achieved; however, this reduces the impedance of the slot, $Z_{slot}$. In this case, the asymmetric stripline can be used to increase $Z_{slot}$, as proposed in [18]. Decreasing the distance between the strip and the slot, $h_1$, while keeping constant $h$ (increasing $h_2$) leads to higher values of $Z_{slot}$. However, if $h_1$ is too low, the strip will interfere with the field distribution of the slot mode and the lattice network will not be able to separate the TEM and slot modes properly. Finally, condition (2b) can be fulfilled by adjusting the length of the strip. This length only modifies the electrical length of the transmission line in the model shown in Fig. 5(a), that is, the position of the zero of $Z_b$.

\section{Implementation and Results}

\begin{figure}[t!]
\centering
\centerline{\includegraphics[width=0.85\columnwidth]{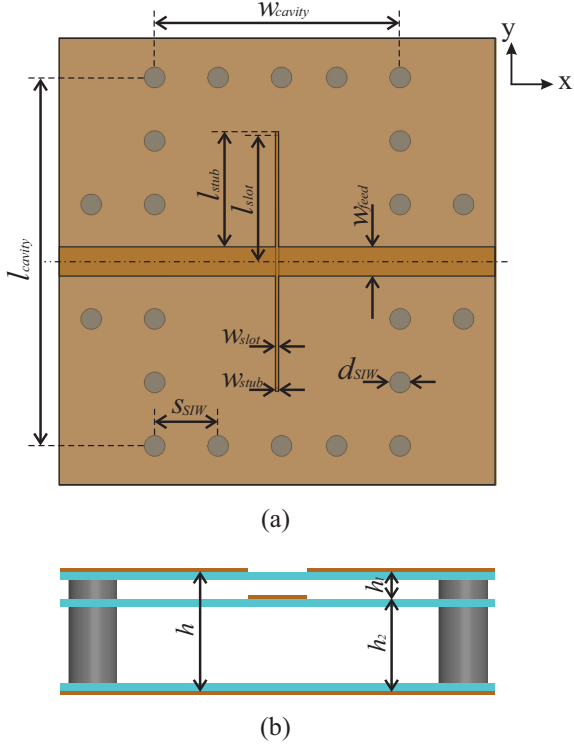}}
\caption{Simulated proof of concept structure. (a) Top view. (b) Section of the cavity. The stub and slot sections have been enlarged to improve the visibility of the elements. }
\end{figure}

\begin{figure}[t!]
\centering
\centerline{\includegraphics[width=0.9\columnwidth]{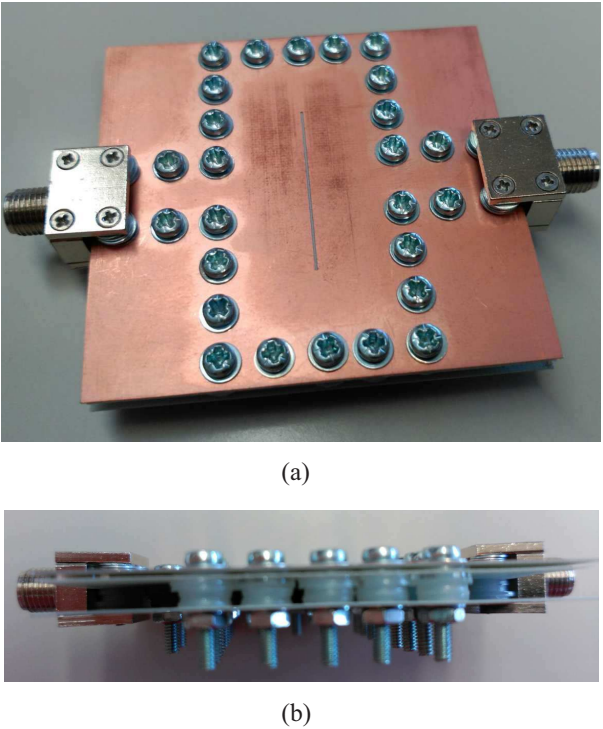}}
\caption{Manufactured prototype. (a) Top view ($w_{feed}$ = 2.8~mm, $w_{cavity}$ = 24~mm, $l_{cavity}$ = 35~mm, $w_{stub}$ = 0.3~mm, $l_{stub}$ = 11.1~mm, $w_{slot}$ = 0.3~mm, $l_{slot}$ = 12.15~mm, $h$ = 3.912~mm, $h_1$ = 0.787~mm, $h_2$ = 3.125~mm, $d_{SIW}$ = 2~mm and $s_{SIW}$ = 6.1~mm). (b) Lateral view.}
\end{figure}

In order to make the overall size of the structure larger, easing the manufacturing requirements, a low permittivity substrate was chosen to implement a proof of concept structure. Thus, a suspended stripline configuration has been chosen, which also allows for a higher flexibility in the implementation of the heights $h_1$ and $h_2$. Three metallic layers have been printed on commercial substrates: lower ground plane, upper ground plane with the slot and the stripline layer. To suspend them in the air, nylon washers have been used around steel screws which pierce through the three layers as metallic posts to make the lateral walls of the cavity. Fig. 6 shows the geometry of the proof of concept structure.

The substrate used to manufacture the three metallic layers is Rogers RO4350B with thickness of 0.25 mm and $\varepsilon_r$ of 3.66. The width of the feeding stripline, $w_{feed}$, was chosen to have a characteristic impedance, $Z_0$, of 50 $\Omega$. The width and height of the cavity, $w_{cavity}$ and $h$, and the width of the slot, $w_{slot}$, were chosen to ensure that the slot mode propagates from 3.5 GHz and the cut-off of the TE$_{10}$ mode is 6 GHz. The width of the stub, $w_{stub}$, together with the asymmetry of the stripline ($h_1$ and $h_2$) were selected to achieve condition (2b). With the current implementation possibilities, it would not have been possible to satisfy this condition using a higher effective dielectric constant. The lengths of the complementary elements, $l_{slot}$ and $l_{stub}$, were designed to ensure complementarity, fulfilling condition (2a). The pole of $Z_a$ and the zero of $Z_b$ are placed at 5.2 GHz. The length of the cavity, $l_{cavity}$ is chosen to be high enough not to affect the behavior of the strip and slot ends. The chosen value results in an approximate resonance of the TE$_{101}$ resonant mode at 6.7 GHz. The analysis in previous sections did not take into account either the multiple layer structure or the presence of the metallic posts and nylon washers, so some tuning of the dimensions using HFSS was needed to obtain the final design. The prototype was manufactured and assembled. 50 $\Omega$ standard stripline connectors were used. Fig. 7 shows the result of the manufacturing process and its dimensions.

\begin{figure}[t!]
\centering
\centerline{\includegraphics[width=0.9\columnwidth]{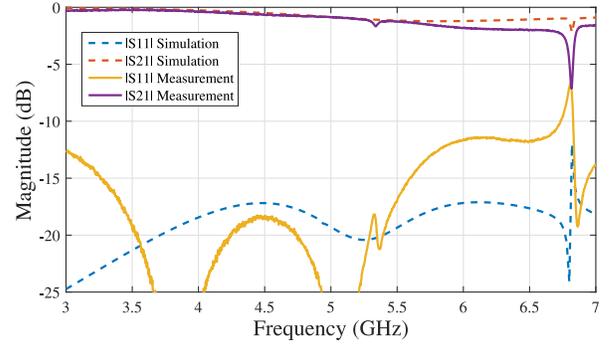}}
\caption{Magnitude of the S-parameters of the simulated and measured structure.}
\end{figure}

In order to place the reference planes at the center of the element (coincident), a TRL calibration kit was designed. Fig. 8 shows the magnitude of the $S_{11}$ and $S_{21}$ parameters for both the simulated and measured cases. Using the criterion of -10 dB for the $S_{11}$ to determine the impedance bandwidth, impedance matching of up to 6.7 GHz is foundin the measurements, versus 6.8 GHz in the simulation. The CBS does not exhibit significant radiation below 4 GHz (radiated power of less than 5\% of the input power), which limits the use of this radiating element at lower frequencies and, thus, a fractional bandwidth of about 48\% is obtained (50\% in the simulation). The spurious ripple at 6.8 GHz can be explained due to the appearance of the TE$_{101}$ resonant mode, as was expected from the analysis of Section II. It can be shown analytically that a small difference in either the lengths of the stubs or the effective dielectric constants can produce a kind of ripple at the frequency corresponding to the zero of $Z_b$. Therefore, the small ripple found in measurements around 5.3 GHz is attributed to a difference in the electrical length of both stubs, due to manufacturing errors.

\begin{figure}[t!]
\centering
\centerline{\includegraphics[width=\columnwidth]{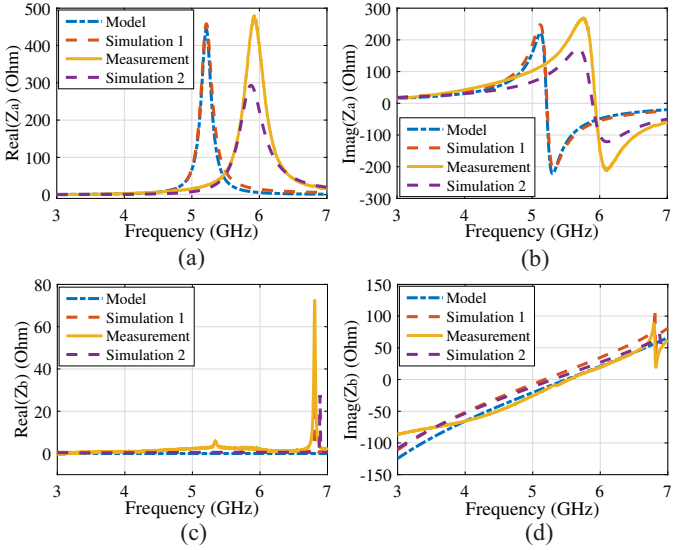}}
\caption{Lattice network impedances of the radiating element: model, design simulation (with nominal dimensions), prototype measurement and the prototype simulation (with actual manufactured prototype dimensions). (a) Real part of $Z_a$. (b) Imaginary part of $Z_a$. (c) Real part of $Z_b$. (d) Imaginary part of $Z_b$.}
\end{figure}

To explain the differences between the model, simulation and measurements in Fig. 8, the values of the impedances of the lattice network, $Z_a$ and $Z_b$, are extracted from the S-parameters and compared with those of the transmission line model of Fig. 5. The results are shown in Fig. 9. Good agreement between the model and simulation is found, which proves the validity of the independent designs of the stub and CBS elements. It can be seen that the pole of $Z_a$ and the zero of $Z_b$ appear around the design frequency, 5.2 GHz, resulting in the complementarity of the structure. The frequency shift in the pole of $Z_a$ in the measured case explains the difference between the simulation results and the measurements in Fig. 8. Due to limitations of the manufacturing technology available, the thicknesses of the substrates are thinner than expected and they bend a little across their surfaces, especially on the slot, where there is almost no dielectric left. The absence of dielectric near the slot reduces the effective $\varepsilon_r$ of the slot mode, leading to a frequency shift. In order to verify this effect, additional simulation results using the actual dimensions of the prototype have been included in Fig. 9 as \textit{Simulation 2}. The missing dielectric in the center of the slot has been simulated with a length of 10 mm, and the thicknesses of the top, middle and bottom layers, instead of being the target 0.25 mm, are around 0.15 mm, 0.2 mm and 0.24 mm respectively. These changes explain the frequency shift. Given the sensitivity of the element to the thicknesses of the substrates and the uncertainties introduced by the manufacturing process, the discrepancies in the impedance level are reasonable. The effect of the bending and reduced thicknesses of the substrates is also present in the manufactured TRL calibration kit and may have introduced additional errors in the measurements.

\begin{figure}[t!]
\centering
\centerline{\includegraphics[width=0.75\columnwidth]{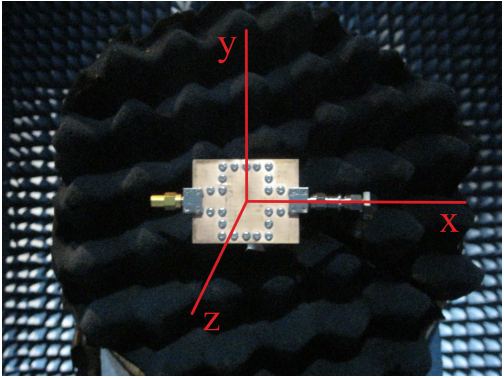}}
\caption{Proof of concept structure built and measured. Radiation measurements made in the anechoic chamber of the Laboratorio de Ensayos y Homologación de Antenas, Universidad Politécnica de Madrid, Madrid (Spain).}
\end{figure}

\begin{figure}[t!]
\centering
\centerline{\includegraphics[width=0.95\columnwidth]{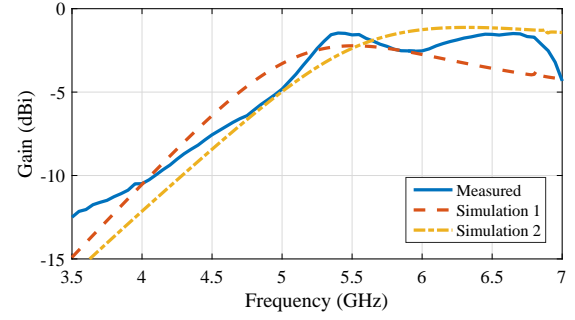}}
\caption{Measured and simulated gain of the radiating element sampled in broadside direction.}
\end{figure}

\begin{figure}[t!]
\centering
\centerline{\includegraphics[width=\columnwidth]{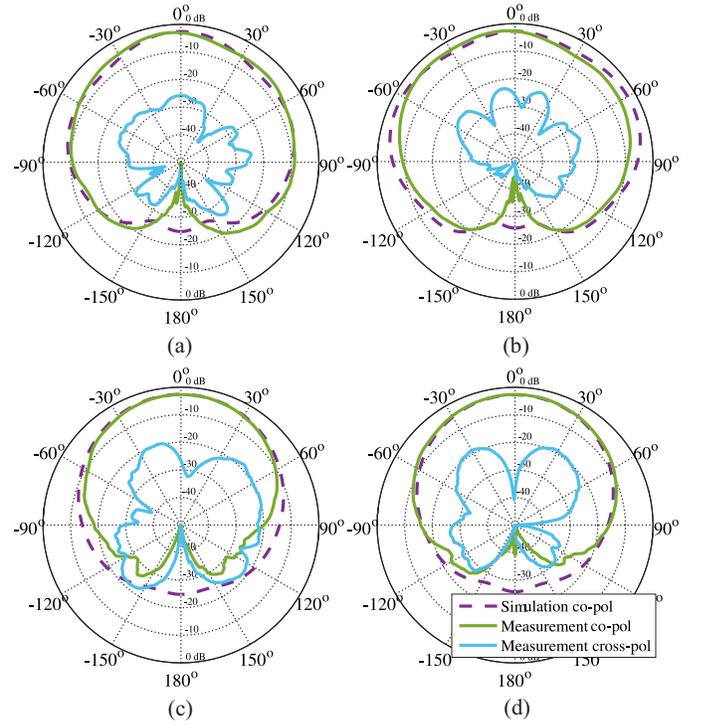}}
\caption{Radiation gain patterns of the simulated and measured structure. (a) 5.2 GHz, XZ plane. (b) 6 GHz, XZ plane. (c) 5.2 GHz, YZ plane. (d) 6 GHz, YZ plane.}
\end{figure}

Finally, the radiation properties of the prototype have been measured, as shown in Fig. 10. Fig. 11 represents the gain over frequency for both the simulated cases, and the measurement in the broadside direction (+Z-axis in Fig. 10). Given the sensitivity of the manufacturing process, reasonable agreement is found between the measurements and the simulations, showing an increasing gain up to the frequency of 5.5 GHz, from which it slowly decreases. Some discrepancies were expected as there are significant differences between the simulated gains in the structure with the target design dimensions and the structure with the manufactured actual dimensions. The low values of gain are expected, since the power radiated by this two-port element is only about 20\% or less of the input power throughout the working band. This is due to power leaking to the second port as desired, since the structure is proposed as a radiating element for series-fed arrays. Fig. 12 shows the radiation gain patterns of the simulation, using HFSS, and manufactured structure at 5.2 GHz and 6 GHz. It can be seen, as expected, that the structure exhibits broadside radiation in only one half-space. Measurement and simulation differences may be due to the presence of the connectors and the heads of the screws, not considered in the simulation. Nevertheless, good agreement between them is found. The polarization obtained is linear in the principal planes with very good cross-polarization discrimination. The cross-polarization simulation results were omitted since they were less than -50 dB.

\section{Use in Series-fed Arrays}

The transmission configuration of this element makes it especially suitable for traveling-wave, series-fed arrays. Its broad impedance matching allows frequency reconfigurable arrays to be built without requiring the modification of the geometry of the element (as needed when resonant elements are used), simplifying the design, as done in [8]. In these arrays, the control of the power radiated by the elements is very important. The power radiated by the proposed element can be controlled by changing the width of the slot. However, as stated in Section II, an increase in the width of the slot leads to a higher cut-off frequency in the CBS mode. To solve this problem, increasing the height of the cavity can help to maintain a wide, working bandwidth. In return, the impedance of the slot mode will change and, in order to satisfy condition (2b), the width of the stub should be readjusted.

\begin{figure}[t!]
\centering
\centerline{\includegraphics[width=0.9\columnwidth]{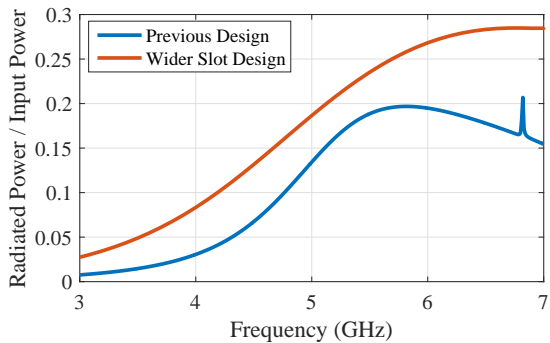}}
\caption{Simulation results of fraction of radiated power of input power of the previous design and a design featuring a wider slot.}
\end{figure}

To illustrate the control of the radiation properties by changing the width of the slot, another design has been simulated. In this case, the width ($w_{slot}$) and length ($l_{slot}$) of the slot are set to 1.5~mm and 13.2~mm respectively. The width ($w_{stub}$) and length ($l_{stub}$) of the stub are 0.8~mm and 10.7~mm respectively. The distance between the strip and the bottom ground plane, $h_2$, has been increased to 5.5~mm. The other parameters have not been modified. Fig. 13 shows the simulated fraction of the input power that is radiated by this element compared to that of the previous design. A significant increase in the percentage of the radiation power can be observed throughout the working bandwidth. This means that a series-fed array with 8 elements radiating each 25\% of their input power would radiate around 90\% of the input power of the array, ignoring losses. This way, by controlling the radiated power of the radiating element, it is possible to choose the size of the array and, thus, the directivity.

The spacing between the elements of the array is also another important parameter to take into account. The minimum distance between elements is limited by the width of the cavity, $w_{cavity}$, since the same row of metallic posts can be used to make the vertical walls of the cavity of two adjacent elements. This space should be enough for most arrays but, if this were not the case, a material with a higher $\varepsilon_r$ could be used if the implementation technology allows it.

\section{Conclusion}

A broadband CBS radiating element fed by stripline with a transmission configuration is proposed. Its main novelty is the enhanced impedance bandwidth obtained when a complementary stub is placed beneath the slot. Modes propagating through the cavity play a fundamental role in its behavior. In order to understand the performance of the structure and simplify its design, a lattice network-based transmission line model was proposed. This methodology greatly avoids the use of a parametric analysis. A proof of concept structure was designed and manufactured, achieving satisfactory simulation and measurement results: S$_{11}$ shows a wide impedance matching, a measured fractional bandwidth of 48\%, and the element exhibits unidirectional radiation in the working bandwidth with a very pure linear polarization. The proposed structure can be used to build reconfigurable wideband, series-fed antenna arrays to be used in bandwidth-intensive wireless communications systems. Future work will focus on finding a more robust implementation of the structure.

\section*{Acknowledgment}

The authors would like to thank the anonymous reviewers for their valuable comments and suggestions to improve the quality of the paper. They are also grateful to Prof. J. Esteban, from Universidad Politécnica de Madrid (Spain), for greatly helping with the understanding of the structure.

\end{document}